\documentclass[%
reprint,
superscriptaddress,
 amsmath,amssymb,
 aps,longbibliography,
 onecolumn,
]{revtex4-2}

\usepackage{esvect}
\usepackage{graphicx}
\graphicspath{ {./figures/} }
\usepackage{dcolumn}
\usepackage{bm}
\usepackage{hyperref}
\usepackage[noabbrev]{cleveref}
\usepackage[dvipsnames]{xcolor}
\usepackage{gensymb}

\begin{document}

\preprint{APS/123-QED}

\title{Mitigation of the Turbulence within an Arteriovenous Fistula with a Stent Implantation}

\author{Sanjiv Gunasekera}
\email{sanjiv.gunasekera@rmit.edu.au}
\affiliation{%
 School of Mechanical of Manufacturing Engineering,
 The University of New South Wales, 2052, NSW, Australia
}%
\author{Tracie Barber}%
\author{Olivia Ng}%
\affiliation{%
 School of Mechanical of Manufacturing Engineering,
 The University of New South Wales, 2052, NSW, Australia
}%
\author{Shannon Thomas}
\author{Ramon Varcoe}
\affiliation{
 Department of Vascular Surgery,
 Prince of Wales Hospital, Randwick, 2031, NSW, Australia
}%
\author{Charitha de Silva}%
\affiliation{%
 School of Mechanical of Manufacturing Engineering,
 The University of New South Wales, 2052, NSW, Australia
}%




\textbf{The revised manuscript has been published in the Physical Review Fluids \href{https://doi.org/10.1103/PhysRevFluids.7.123101}{available here} or at \url{https://doi.org/10.1103/PhysRevFluids.7.123101}}



\begin{abstract}
The transitional flow which initiates within the junction (anastomosis) of an arteriovenous fistula (AVF) is known to be a contributing factor in the onset of vascular disease. A novel treatment method involving the implantation of a flexible stent across the anastomosis has enabled the retention of a large proportion of functioning AVFs, despite the propensity for stent malapposition to occur at the sharp inner curve of the anastomosis. Large Eddy Simulations (LES) of a patient-specific AVF with and without the presence of a stent captured oscillatory flow behaviour emanating from the interface of the two inlet flows in the stent-absent case, however, these oscillatory features were subdued in the stented case. The stent-absent case generally had higher Turbulent Kinetic Energy (TKE) in the anastomosis which led to larger cycle-to-cycle variations in wall shear stress (WSS). The significantly lower TKE generated at the heel of the stented AVF was contained within the malapposed stent, thereby resulting in lower WSS fluctuations. However, a slight increase in turbulence downstream of the malapposed stent edge was noted. This detailed study reveals a significant decrease in turbulence within the AVF in the presence of the stent, thereby, providing a level of understanding underpinning the success of the treatment strategy from a fluid dynamic perspective.
\end{abstract}

\maketitle


\section{\label{sec:stent_intro}Introduction}

Chronic Kidney Disease is the progressive decrease of kidney filtration ability with the last stage of this condition known as End Stage Renal Disease (ESRD). At this stage, the majority of patients that are not granted kidney transplants opt for haemodialysis where the blood is filtered outside the body in a dialysing machine. This option requires vascular access which is usually achieved by creating an arteriovenous fistula (AVF) \citep{santoro2014vascular}, which is a surgical connection created between an artery and vein in the forearm. However, the AVF frequently suffers from Intimal Hyperplasia (IH) where smooth muscle cells propagate and proliferate in the innermost vessel wall layer leading to the narrowing of the blood vessel \citep{duque2017dialysis}. Haemodynamic forces, particularly originating from disturbed flow \citep{gimbrone2016endothelial}, cause injury to the endothelial cells lining the vessel wall and are known to be one of the many pathways that initiate IH \citep{roy2006hemodialysis}. The initiation of transitional flow behaviour at the anastomosis (junction) of the vasculature has been observed in CFD simulations of AVF models \citep{fulker2018high,Bozzetto2016, stella2019assessing} noting the gradual dissipation of the turbulent energy as the flow traverses downstream along the vein.

Treatment of AVF disease can involve either surgical salvation where the anastomosis site is relocated or endovascular options such as balloon angioplasty and stent implantation \citep{bharat2010assessment, tessitore2006endovascular}. Stents, which are metallic mesh-like scaffolding devices that hold open vessels, have been used sparingly in AVFs due to the concerns associated with the hindering cannulation \citep{mallik2011operative} and unintended migration \citep{anwar2017stent}. However, recent use of nitinol stents, known for their elasticity thereby increasing the shape conformity to tortuous vessels \citep{stoeckel2004self}, have yielded high unassisted patencies \citep{swinnen2015juxta, thomas2019interwoven, thomas2022long} retaining a large proportion of functioning AVFs.
 
A side-effect of implanting the flexible stent across the sharp-angled anastomosis is the malapposition of the stent (sometimes referred to as the incomplete stent apposition), where the device is not completely embedded against the wall of the vessel. An undersized stent implantation is known to be a predictor of stent thrombogenicity \citep{van2009predictors} and the malapposition of the stent has been shown to produce recirculation regions near stent struts, along with the increase in areas of low Time-Averaged Wall Shear Stress (TAWSS) \citep{chen2017haemodynamic, beier2016hemodynamics}. However, with the increase in the degree of undersizing beyond 5\% of the vessel diameter (equivalent to a decrease in stent lumen area below 90.25\% of the vessel lumen) a decrease in the heterogeneity of the wall shear stress (WSS) was evident across the vessel \citep{chen2009effects}. Flow simulations of patient-specific models have also observed a lower coverage of adverse TAWSS in the presence of a malapposed stent \citep{wei2021impact}. A recent study has shown that undersized stent-grafts used to treat cephalic arch stenoses associated with AVFs, have yielded higher access patency rates and a decrease in interventions \citep{huang2020undersized}.

Comparison of the AVF flow field with and without the flexible stent implantation highlighted the significant funnelling of flow within the malapposed stent encapsulated region, potentially containing the adverse haemodynamic behaviour away from the vessel wall \citep{gunasekera2021impact}. This funnelling characteristic is commonly leveraged in intracranial vasculature to divert flow away from aneurysms \citep{brinjikji2013endovascular}. The diversion of a significant portion of flow away from the aneurysm leads to occlusion in most cases \citep{kulcsar2012flow} which is a better outcome than the converse realistic possibility of aneurysm rupture.

The present study aims to characterise the turbulent behaviour of AVF flow in the presence of the stent, unravelling the interaction between the bulk flow funnelling feature of the stent and the inherently turbulent AVF anastomosis. To this end, the turbulent flow features are resolved with Large Eddy Simulations (LES) of a single patient-specific AVF with and without the presence of a stent implantation. The same vessel geometry is used to isolate the influence of the stent. Comparisons of the fluctuating bulk flow field are followed by the assessment of the cycle-to-cycle variations in WSS.
 
\section{\label{sec:stent_method}Methodology}

\subsection{Patient data retrieval}

The AVF of a 78-year-old male volunteer (at rest) was scanned with informed consent under the approval of the Low/Negligible Risk (LNR) application with the South Eastern Sydney Local Health District Human Research Ethics Committee number 15/063 (LNR/16/POWH/7). A non-invasive three-dimensional freehand ultrasound scanning method was used to obtain the geometry of the AVF \citep{colley2018methodology}. The ultrasound images were synchronised with the location of the ultrasound probe using an in-house MATLAB (The Mathworks Inc.) code, to create a volume of scans. Subsequently, manual segmentation along with 3D Gaussian smoothing functions, conducted in Simpleware ScanIP image processing software (Synopsys Inc., CA, USA), resulted in the reproduction of a surface mesh file (in the STL format) of the vessel lumen.

\begin{figure*}[ht!]
\includegraphics[width=15cm]{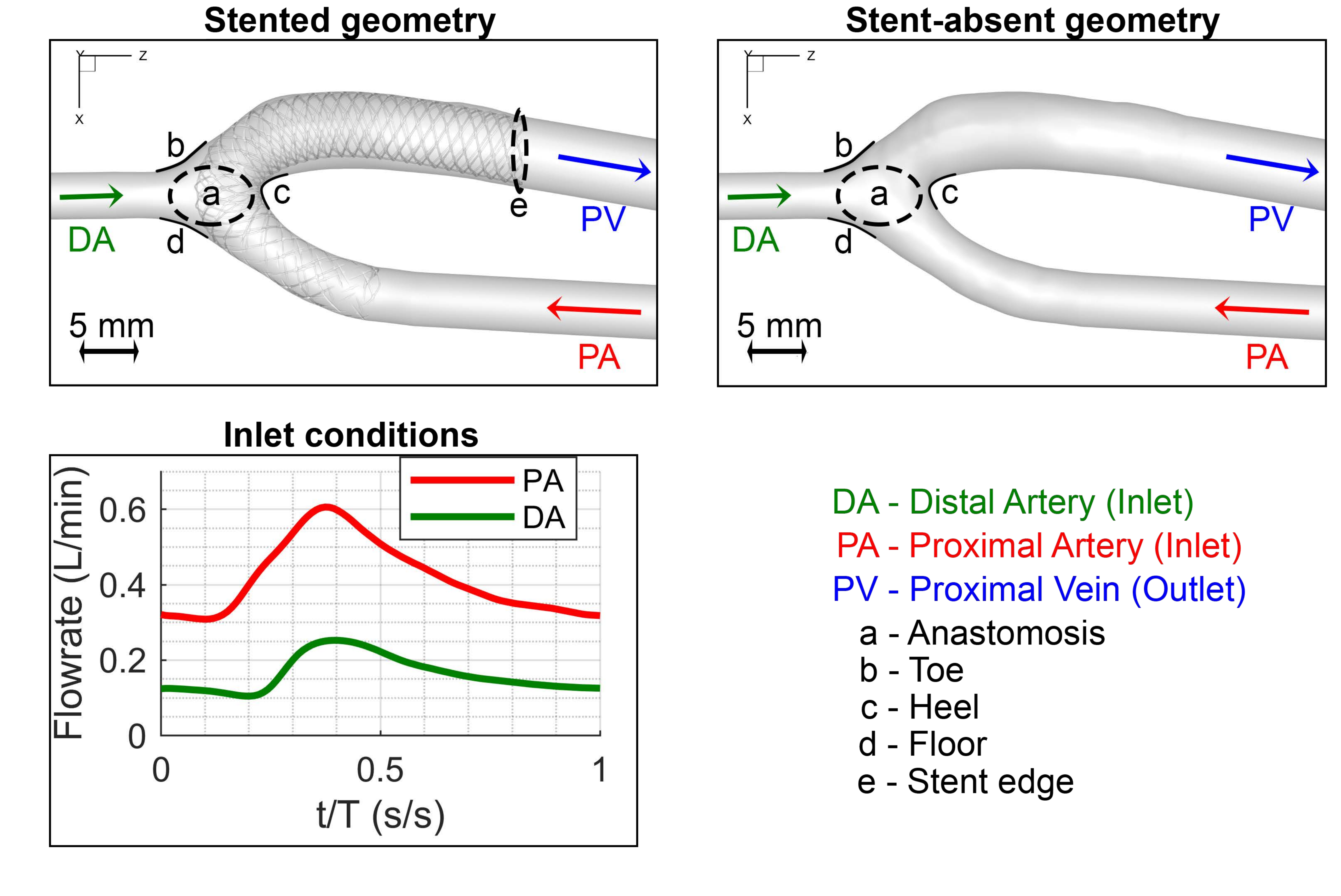}
\caption{\textbf{a)} A segment of a stent strut with the centerline imposed across the strut. \textbf{b)} The physiological inlet flow waveforms imposed on the proximal artery (red) and distal artery boundaries (green). \textbf{c)} The AVF geometry with the stent implantation and annotations of important regions.}
\label{fig:afmc_overview}       
\end{figure*}

The initial ultrasound scans could not be used to segment the stent due to the presence of excessive artefacts and the inherent low resolution of the scanning methodology. To capture relatively higher resolution images of the stent using a micro-CT scanner a polydimethylsiloxane (PDMS) phantom of the AVF geometry was fabricated \citep{gunasekera2020numerical}. PDMS was cast around a water-soluble 3D print of the AVF geometry which was eventually flushed out. To obtain a precise rendering of the stented AVF model, a micro-CT scanner (MILabs U-CT at MWAC, UNSW Sydney) with a step angle of $0.25\degree$, was used to obtain an ultra-focus scan at an approximate overall resolution of $40~\mu m$. The resulting high-resolution scan volume was segmented to generate an STL of the stent geometry and the segmentation of the vessel geometry was repeated to locate the AVF vessels in the same local coordinate system of the stent STL as shown in figure \ref{fig:afmc_overview}.

The stent and vessel geometries were smoothed using several iterations of a Gaussian smoothing algorithm. To ensure the stent strut diameter was equivalent to $152~\mu m$ (the physical stent strut diameter), sections of the segmented stent radius were measured using MATLAB (Mathworks Inc.) by calculating the distance between the stent strut centerline and the closest STL surface elements. This method was used after each smoothing iteration to ensure an appropriate level of surface smoothness was achieved whilst avoiding overly shrinking the stent STL.

The spectral doppler function of the ultrasound machine was used to extract the patient-specific flow rate boundary conditions of the inlet vessels \citep{carroll2020tracking}. The patient AVF under consideration had a retrograde inlet flow configuration where the proximal artery and the distal artery supplied blood to the outlet vein of the AVF. Since the inlets of the two vessels were recorded individually, a simultaneous ECG trace was recorded using a 3-lead ECG. The peak of the `QRS' complex of the ECG trace was used to temporally synchronise the inlet waveforms of the two vessels. The Doppler images, which were cycle-averaged and processed using a median filter and Gaussian smoothing, were digitised using a MATLAB algorithm \citep{carroll2020tracking}.

\subsection{Numerical calculation setup}
\label{sec:cfdmethods}

The presence of flow transitioning from the laminar regime, particularly in the deceleration phase of pulsatile flow, within vascular geometries has been noted in direct numerical simulations (DNS) \citep{valen2011direct, varghese2007direct}. Therefore, to understand any haemodynamic disturbances within the AVF, a simulation that is able to resolve any turbulence generation is of value. Since obtaining the spatial and temporal resolution to conduct a DNS is highly resource intensive, a Large Eddy Simulation (LES) that resolves a significant portion of the turbulence generation can provide inferences of the behaviour of larger, higher energy containing eddy disturbances while modelling the energy of the eddies smaller than the grid size. 

To resolve larger scales while modelling smaller scales, an LES makes use of a spatial filter which results in the filtered Navier-Stokes equations as shown below.

\begin{equation}
\label{eq:continuity_filtered}
    \nabla \cdot \tilde{U} = 0 \\
\end{equation}

\begin{equation}
\label{eq:momentum_filtered}
    \rho \bigg(\frac{d\tilde{U}}{dt}+(\tilde{U} \cdot \nabla)\tilde{U} \bigg) = -\nabla \tilde{P} + \mu \nabla^2\tilde{U} - \nabla \cdot \tau_{SGS} \\
\end{equation}

The filtered velocity and pressure variables are denoted by $\tilde{U}$ and $\tilde{P}$, respectively. To close the filtered Navier-Stokes equations, values for the subgrid scale turbulent stresses, denoted by $\tau_{SGS}$, are modelled. The dynamic Smagorinsky subgrid scale (SGS) model was chosen to capture the lower energy disturbances and contribute to the dissipation of turbulent energy. The original Smagorinsky SGS model is based on the assumption that small scale turbulent energy (contained within the grid) is isotropic. The Smagorinsky constant ($C_S$) is used to estimate the subgrid energy which is proportional to the grid length scale. Therefore, the model had a tendency to be overly dissipative at regions close to the wall. Improvements to the model, such as the use of a mixing length (instead of the grid length alone) and a dynamically varying $C_S$ value based on an additional filter length, twice the size of the grid length, have made the model more robust in applications of wall bounded flow \citep{germano1991dynamic,lilly1992proposed}. However, the subgrid viscosity does not go to zero in laminar shear flow regions, thereby still maintaining the behaviour of being overly dissipative in regions of separating flow \citep{nicoud1999subgrid, nicoud2011using}. Despite this drawback of the SGS model, studies of stenotic flow have shown that the model replicates the mean flow and turbulent statistics well when compared with results from DNS \citep{jabir2016numerical, pal2014large, tan2011comparison}. Similarly, a patient-specific study of a stenosed region in a carotid artery demonstrated that LES with the dynamic Smagorinsky model was able to replicate the mean flow and turbulent features with pulsatile inlet conditions \citep{mancini2019high}, thereby showing the applicability of the SGS model in the current AVF study as well.

The fluid domain was initialised using the $k-\omega~SST$ turbulence model based simulation that has been detailed in previous studies \citep{carroll2020tracking, gunasekera2020numerical, gunasekera2021impact, ng2022effect}. The fluid region of the AVF geometry was discretized by creating a tetrahedral mesh in the bulk flow and boundary layer regions using ICEM CFD (ANSYS Inc.). Tetrahedral elements were used, even in the boundary layer region, because of the complexity of the geometry which incorporated the small intertwining surface of the stent within the relatively larger vessel surface. Approximately 90\% of the elements were of high aspect ratio and low skew. The grid convergence index (GCI) \citep{celik2008procedure} was calculated by using the average of the velocity magnitude across a vessel lumen slice in the vein. The average GCI values were 0.016\% and 1.95\% for the stented and stent-absent geometries, respectively. Additionally, circumferentially averaged wall shear stress (WSS) values along the vein were compared against those from the coarse and fine mesh. The WSS values were largely seen to match the values of the fine mesh in both geometries with an average relative error of 1.86\% and 1.63\% for the stented and stent-absent geometries, respectively. Therefore a sufficient level of spatial discretization was achieved with 19.75 million elements for the stented geometry and 4.89 million elements for the stent-absent geometry to conduct the initial Reynolds-averaged Navier-Stokes (RANS) simulations.

\begin{figure*}[h]
\includegraphics[width=15cm]{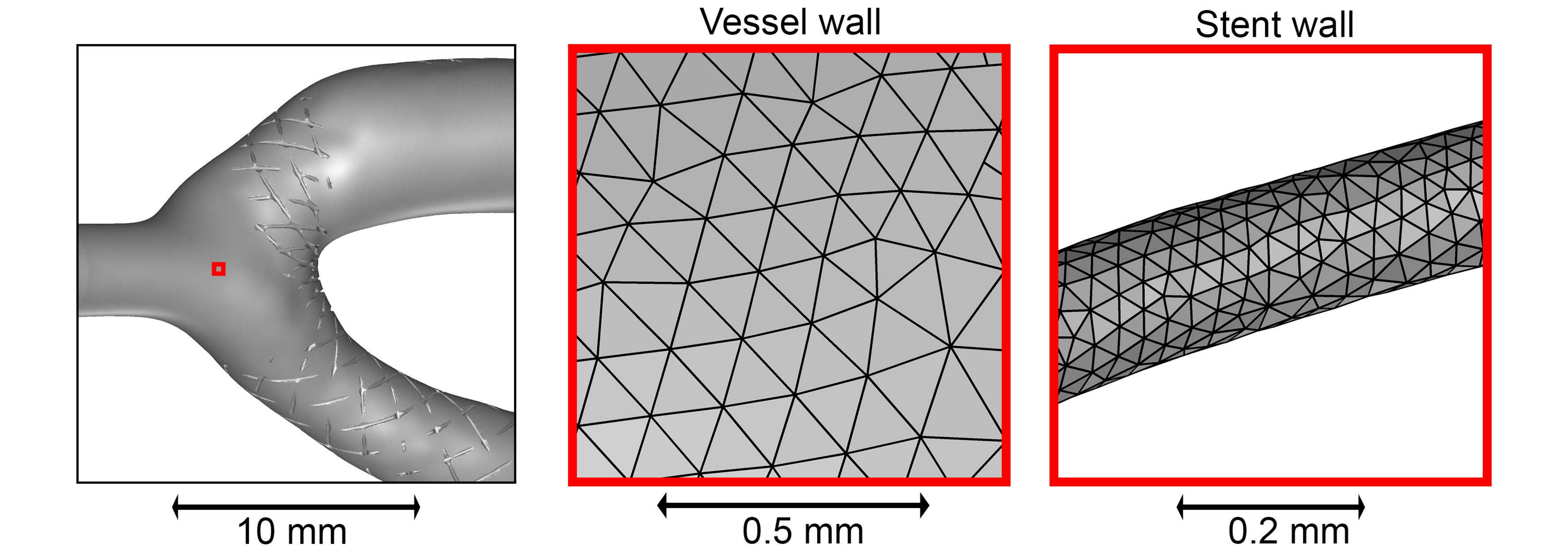}
\caption{Detail of cells at the vessel and stent wall boundaries near the anastomosis.}
\label{fig:wall_mesh}       
\end{figure*}

The mesh was further refined using a mesh adaption algorithm which refined elements, with a value of the gradient of subgrid dynamic viscosity, greater than a stipulated value. The chosen value was altered such that $1-5\%$ of the total cells were marked at each mesh adaption step. Basing the mesh adaption on the gradient of the subgrid viscosity led to marking regions where sudden changes in dissipative scales were present (such as in the anastomotic region and the along the vein) thereby providing the opportunity to further resolve these regions to avoid overly dissipating the turbulence generated in important regions. Four mesh adaption steps were conducted (two each at the maximum and minimum point of the cycle) to refine a sufficient number of elements to capture behaviour across the cycle. Application of this targetted mesh refinement methodology resulted in meshes composed of 24.60 million (increasing from 19.75 million) and 6.91 million (increasing from 4.89 million) elements for the stented and stent-absent cases. The average cell edge length on the stent wall was $32.5~\mu m$ which resulted in approximately 17 cells back-to-back across the circumference of a $152~\mu m$ stent strut. The average cell edge length on the vessel wall was $103~\mu m$ which resulted in approximately 176 cells back-to-back across the circumference of a vessel segment that is $5~mm$ in diameter. Using the stent strut diameter and the average outlet velocity (as the malapposed stent struts faced high velocities), flow with a Reynolds number of 25 would be the maximum faced at the stent struts. The detail of the mesh at the stent wall and vessel wall have been illustrated in figure \ref{fig:wall_mesh}.

The pulsatile patient-specific inlet flow waveforms were applied at the proximal artery and the distal artery boundaries of the AVF geometry. The recorded heart rate of the patient was 60 bpm which was matched in the simulations by setting the period of the inlet waveforms to $1~s$. The proximal vein was maintained as a non-tractive outlet condition with relative pressure held at zero. To ensure the inlet flow waveforms had a developed flow profile, the ends of the vessels were extended by 10 times the diameters of the relevant vessel. The initial transient RANS was conducted with a fixed time-step of $0.001~s$. Comparison of the slice-averaged velocity magnitude values between a time-step of $0.001~s$ and $0.0001~s$ showed a difference of 0.0218\% and 0.166\% for the stented and stent-absent cases, respectively. The difference in the circumferentially averaged WSS measurements were 1.50\% and 1.52\%, respectively, therefore the time-step of $0.001~s$ was deemed suitable. With the increased resolution of the spatial discretization for the LES, the temporal discretization required refinement as well to maintain the Courant-Friedrich-Lewy (CFL) (defined in equation \ref{eq:cfl}) below 1, such that the distance travelled by the flow within a time-step, remains within the bounds of the cell \citep{courant1967partial}.

\begin{equation}
\label{eq:cfl}
    CFL = \frac{U\Delta t}{\Delta x}\\
\end{equation}

The velocity magnitude, time-step and grid length are denoted by $U$, $\Delta~t$ and $\Delta~x$, respectively. Although the Fluent solver employs an implicit solver (which remains stable for Courant numbers larger than 1), it is recommended that this aforementioned condition is maintained to accurately resolve the LES temporally \citep{menter2012best}. After decreasing the time-step to $0.0002~s$ (from $0.001~s$), the average Courant numbers across the domain at the maximum inlet time points were 0.45 and 0.51 for the stented and stent-absent models, respectively. 

The vessel and stent walls were set up with a no-slip wall condition and without compliance. A comparison between a stationary wall to a compliant wall in AVF flow simulations has shown that the resulting WSS trends remain similar with a general overestimation of the values at regions of low WSS in the case of the stationary wall \citep{mcgah2014effects}. The fluid properties were set to replicate the average blood physical properties with a density of $1060~kg/m^{3}$. The effect of the non-Newtonian nature of blood, known to be minimal at the high shear rates \citep{vijayaratnam2015impact} encountered in an AVF, was modelled using the Carreau model \citep{cho1991effects} provided in equation \ref{eq:carreau}. 

\begin{equation}
\label{eq:carreau}
    \mu = \mu_\infty + (\mu_0-\mu_\infty)[1+(\lambda\dot{\gamma})^2]^{\frac{n_c - 1}{2}} \\
\end{equation}
The shear rate variable is denoted by $\dot{\gamma}$, and the model parameters: zero-shear rate viscosity ($\mu_0$), infinite-shear rate viscosity ($\mu_\infty$), characteristic time ($\lambda$) and power-law index ($n_c$) were set to $0.056~Pa\cdot s$, $0.00345~Pa\cdot s$, $3.313~s$ and $0.3568$, respectively.

The pressure and velocity variables were solved using the pressure-based SIMPLE algorithm. A second order upwind scheme was used for the spatial discretization of pressure quantities and a bounded central difference scheme was used for the spatial discretization of momentum quantities. A second order implicit scheme was used for temporal discretization.

It is recommended that an LES is capable of resolving 80\% of the turbulent kinetic energy ($TKE$) \citep{pope2001turbulent}. This requirement was assessed to determine the appropriateness of the level of spatial resolution. The resolved turbulent kinetic energy was calculated using a phase-averaged approach to account for the pulsatile nature of the flow domain. Phase-averaged turbulent kinetic energy was calculated at the maximum and deceleration time points of the cycle using equation \ref{eq:resolvedtke}.

\begin{equation}
\label{eq:resolvedtke}
    TKE_{resolved}=\frac{1}{2}({u'}_{x}^{2}+{u'}_{y}^{2}+{u'}_{z}^{2})\\
\end{equation}

Here, ${u'}_{x}=U_x-\overline{U_x}$, ${u'}_{y}=U_y-\overline{U_y}$, ${u'}_{z}=U_z-\overline{U_z}$ correspond to the instantaneous velocity fluctuations, where $\overline{U_x}$, $\overline{U_y}$, $\overline{U_z}$ are the phase-averaged velocities. In addition to the maximum and minimum time points, $TKE_{resolved}$ was calculated at two points in the acceleration phase and three points in the deceleration phase. This approach was necessary as calculating $TKE_{resolved}$ for all time points would have overwhelmed the storage resources due to the small time-step and grid size. The $TKE_{resolved}$ was phase-averaged over 12 cycles to ensure that the quantity was converged. A user-defined function was employed to update the phase-averaged quantities on each passing cycle in an efficient manner. The unresolved turbulent kinetic energy (contained within the subgrid) was estimated using the following equations,

\begin{equation}
\label{eq:unresolvedtke}
    TKE_{unresolved} = \frac{(C_S\Delta S)^2}{0.3}\\
\end{equation}

\begin{equation}
\label{eq:gridturb_est}
    \Delta = \big(6\sqrt{2}(\text{\textit{cell-volume}})\big)^{1/3}\\
\end{equation}

\begin{equation}
\label{eq:resolvedtke_ratio}
    ratio~of~TKE_{resolved} = \frac{TKE_{resolved}}{TKE_{resolved}+TKE_{unresolved}}\\
\end{equation}

where $\Delta$ is the subgrid length of the tetrahedral elements and $S$ is the strain rate magnitude as calculated within the solver \citep{menter2012best}. The average ratio of $TKE_{resolved}$ across the domain, calculated at the maximum inlet flow time point, was 96.8\% and 97.03\% for the stented and stent-absent models, respectively, thereby showing that the spatial resolution of the mesh was largely sufficient. The average $y+$ from the vessel wall across the domain at the maximum time point was 0.51 and 0.45 for the stented and stent-absent cases. Therefore, the majority of the near-wall cells were placed well within the viscous sub-layer, thereby accurately calculating the wall shear stresses.  

\section{Influence of stent on venous bulk flow behaviour}

\subsection{\label{sec:stent_recirc}Bulk flow oscillations}

The temporally resolved volumetric flow field within the stent-absent geometry was previously captured using Tomo-PIV \citep{gunasekera2020tomographic}. A characteristic feature of the flow within an AVF is the stagnant low velocity flow at the heel of the anastomosis which was captured in the Tomo-PIV measurement. Additionally, the low velocity region was noted to persist across the cycle albeit with slight differences in its spatial extent at the various time-points. Streamlines were utilised to show the presence of a recirculation zone at certain time points in the low velocity region as well. Another low velocity region was noted at the floor of the anastomosis and can be seen throughout the cycle with varying spatial extent, although covering a much smaller spatial extent. This low velocity region moved across the floor of the anastomosis at various cycle time-points, suggesting that the variation of momentum of the two opposing inlet flows across the cardiac profile suppressed each other to varying degrees.

A subsequent study using unsteady RANS included a comparison of the flow behaviour with and without the presence of the stent within the same AVF geometry \citep{gunasekera2021impact}. As with the Tomo-PIV derived flow fields, a low velocity region was noted at the heel of the stent-absent AVF geometry. Although there was a low velocity region in the stented AVF as well, the spatial extent of it was much smaller than that in the stent-absent AVF. Spiralling streamlines were noted to originate from the anastomosis and reattach downstream of the low velocity magnitude location. As expected, the reattachment point was further downstream in the stent-absent case. The high velocity flow tended towards the upper and outer walls of the vessel while the low velocity flow aggregates to the central and lower regions of the vessel, suggestive of vortical flow emanating from the anastomosis region. This flow feature, which has been seen in computational studies of AVFs \cite{fulker2018high}, could be caused due to the flow being directed along a curved trajectory across the anastomosis at a high velocity, and the traditional AVF geometry (with an acute-angled anastomosis) has been seen to further aggravate this feature \cite{lee2016assessing,carroll2019reduction}. 

\begin{figure*}[h!]
\centerline{\includegraphics[width=17cm]{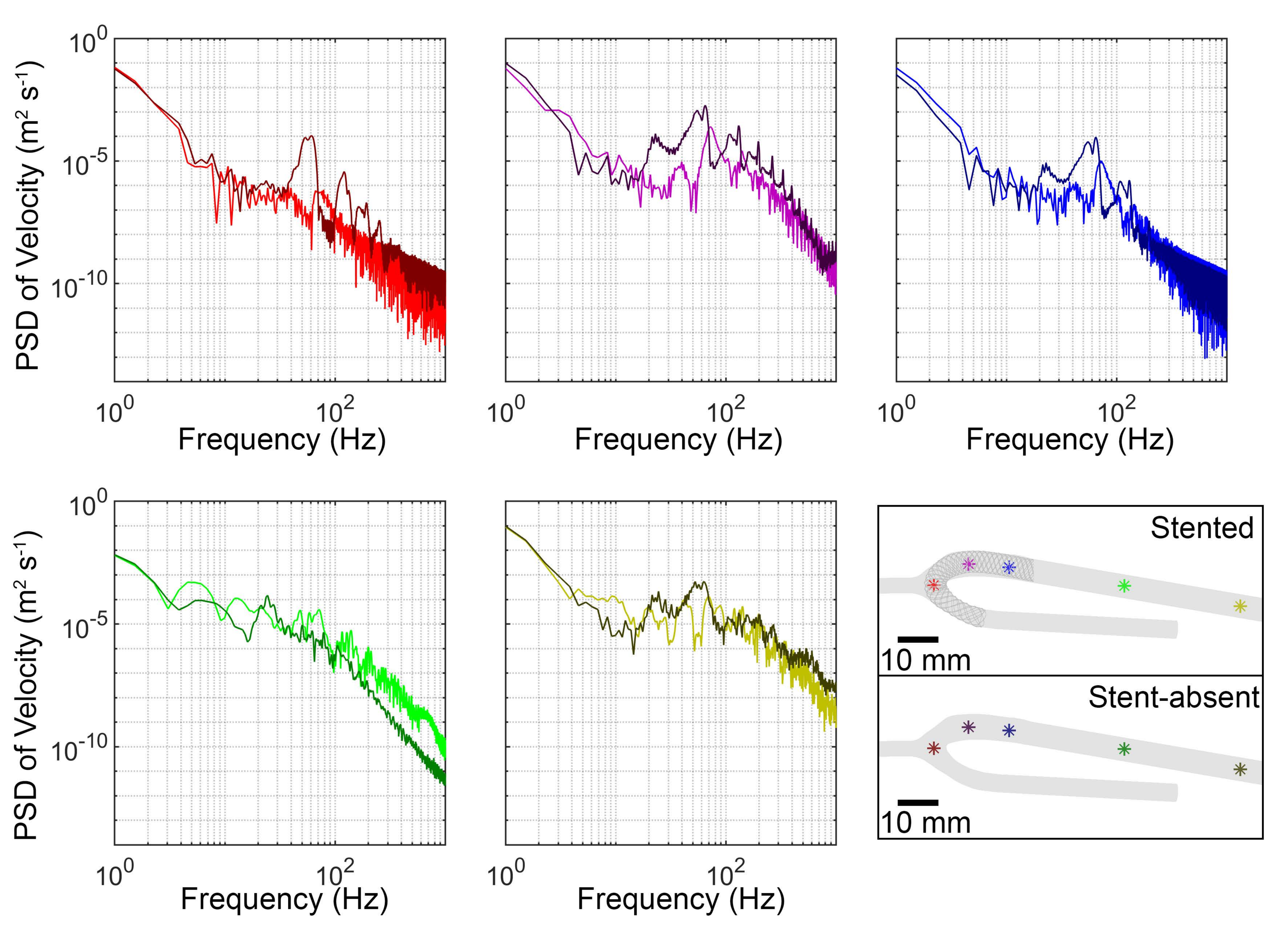}}
\caption{Spectral decomposition of the velocity magnitude time-traces at monitor points along the vein for the stented and stent-absent cases. The darker lines correspond to the stent-absent case while the lighter lines correspond to the stented case. The monitor point locations shown in the sub-figure at the bottom right, link to the spectral plots by colour.}
\label{fig:fft}
\end{figure*}

To further detail the flow behaviour noted in the vein of the AVF, in the current study, five monitor points were placed along the centre of the vein starting from the anastomosis within both the stented and stent-absent geometries as shown in figure \ref{fig:fft}. The velocity magnitude signal recorded at these points was decomposed using Welch's power spectral density (PSD) estimator (within MATLAB R2019b) with a window length equal to the period of a cycle. As expected, the global maximum amplitude for all signals took place at the patient's heart rate with a decrease in amplitude with the increase in frequency. However, a clear local maximum amplitude was noted at frequencies ranging from $50~Hz$ to $70~Hz$ in the stent-absent AVF. This behaviour was predominantly occurring at the first three monitor points (the first row of figure \ref{fig:fft}) which were located at or close to the anastomosis. There were local peaks noted in the signals of these monitor point locations from the stented AVF as well, although the amplitudes of these peaks were much lower. The oscillatory flow behaviour in the stented geometry was noted to occur at frequencies ranging from $70~Hz$ to $90~Hz$. These oscillatory flow events settled as the flow travelled further downstream of the vein in both geometries (as illustrated in the second row of figure \ref{fig:fft}). A lesser significant local peak was noted at the two monitor points furthest from the anastomosis in the stented AVF. However, the power densities at these peaks were comparable with those of the stent-absent AVF signals.

\begin{figure*}[h!]
\centerline{\includegraphics[width=17cm]{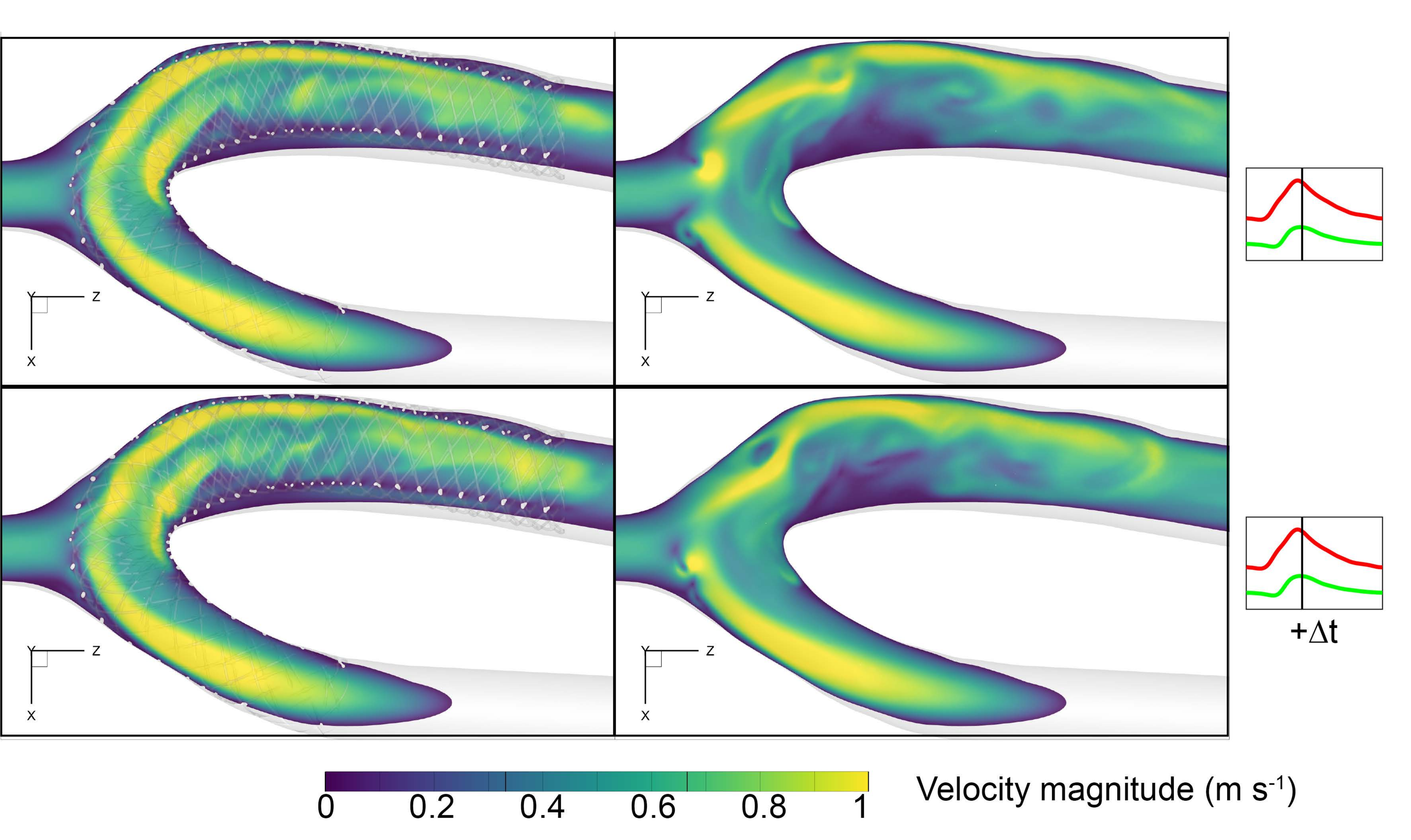}}
\caption{Contour plots of instantaneous velocity magnitude on the central plane of the anastomosis. The distribution is illustrated near the maximum time-point with a time separation $\Delta t = 0.009 s$.}
\label{fig:velmag_amp}
\end{figure*}

These findings suggest that there are significant oscillatory flow features initiating in the anastomosis and persisting through the juxta-anastomotic venous segment in the absence of the stent. Instantaneous velocity magnitude contour plots on the central plane of the anastomosis, shown in figure \ref{fig:velmag_amp}, provide further detail on the aforementioned oscillatory behaviour. The instantaneous flow fields were separated by 0.009 seconds thereby aiming to capture the peaks and troughs of flow oscillations that were occurring at frequencies around $55.56~Hz$. There is a significant change in the spatial location of the high velocity magnitude region at the upper wall of the vein in the stent-absent case. This behaviour can be traced back to the anastomosis where the two opposing inlet flows interact with each other. Similar to the behaviour observed in the Tomo-PIV measurements, the location of the low velocity magnitude region at the floor of the anastomosis has moved noticeably due to the interaction of the inlet flows. In the stented case this interaction takes place closer to the toe of the anastomosis, however, the strength of this behaviour is much less significant with the oscillations drowning out as the flow traverse across the vein. Furthermore, the oscillations occurring in the stent-absent AVF persist throughout the cardiac cycle, however, the oscillations in the stented AVF occur only during the deceleration phase of the cardiac cycle (as seen in the included supplementary video). One reason for the more subdued nature of the inlet flow interaction in the stented AVF is that the distal artery flow is distributed more evenly across the vessel cross-section with the presence of the porous stent however, in the stent-absent AVF, a larger peak is created due to the velocity profile spread across a larger cross-section thereby creating a larger contest with the higher flow coming from the proximal artery.

Another flow behaviour that varies in this short time separation is the low velocity magnitude region at the heel of the stent-absent AVF. The flow that navigates the curved trajectory across the anastomosis eventually separates leading to oscillatory flow behaviour. As before, this behaviour is less pronounced in the stented AVF possibly due to the rough wall surface caused by the texture of the stent surface resulting in a delayed separation.

\subsection{Turbulence generation within the AVF}
\label{sec:stent_tke}

\begin{figure*}[h!]
\centerline{\includegraphics[width=15cm]{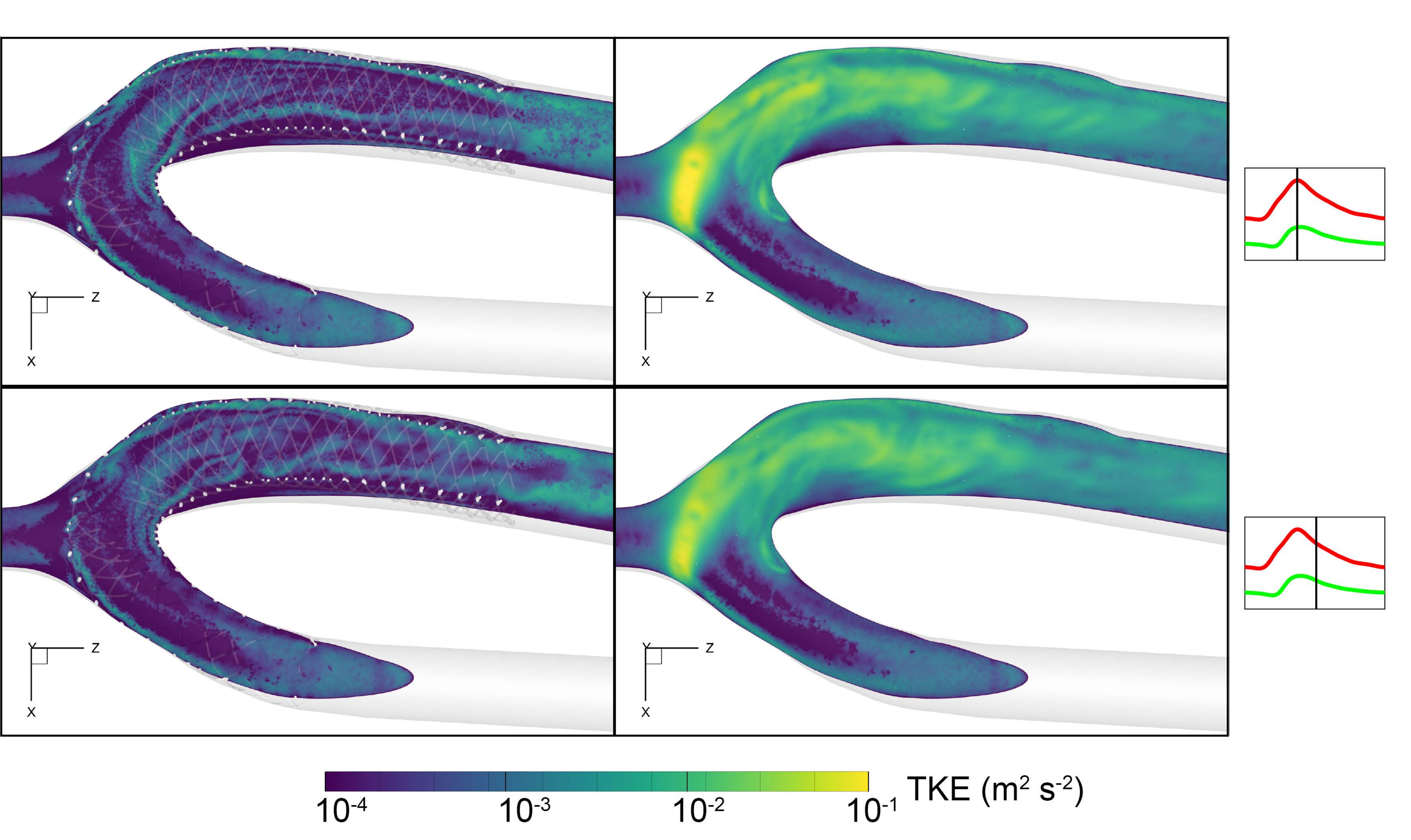}}
\caption{Contour plots of turbulent kinetic energy (TKE) on the central plane of the anastomosis. The distribution of the TKE is illustrated at the maximum and deceleration inlet flow time-points for both stented (left) and stent-absent (right) cases.}
\label{fig:tke_anast_stented}       
\end{figure*}

To quantify the fluctuations generated in the LES of the two AVF cases, $TKE$ was calculated using equation \ref{eq:resolvedtke} that was defined previously. The distribution of the $TKE$ across a central plane cutting through all vessels at the anastomosis is illustrated in figure \ref{fig:tke_anast_stented}. The general level of turbulence generation is much lower (by approximately an order of magnitude) in the stented case when compared to the stent-absent case. 

High $TKE$ persisted throughout the cardiac cycle at the distal region of the anastomosis in the stent-absent case. This region coincides with the oscillations in velocity magnitude noted in the stent-absent case. The variations in velocity from cycle-to-cycle, when there is a slight phase difference in the oscillations, contribute to this increase in turbulence. A parametric study using large eddy simulations of AVF geometries (without stent implantations) of varying anastomosis angles has also revealed a large standard deviation of velocity (proportional to the square root of $TKE$) at the anastomosis \citep{stella2019assessing}. The large regions of variance were located closer to the distal region of the anastomosis, somewhat similar to the findings of the current study. In contrast to the stent-absent case, the $TKE$ generation at the distal region of the anastomosis in the stented case is almost two orders of magnitude lower. The $TKE$ magnitude at the interface of the two inlets is equivalent to the magnitude noted at the walls of the proximal artery and distal artery. Fluctuations larger than those expected in the near-wall regions are absent or insignificant in the stented case (at the interface of the two inlet flows). 

The magnitude of the oscillatory behaviour at the interface of the two inlet flows in the stented case was low, as noted by the spectral decomposition at the first monitor point in figure \ref{fig:fft} and the instantaneous velocity plots of figure \ref{fig:velmag_amp}. Therefore, cycle-to-cycle variations of this behaviour contributed to lower fluctuations in the stented case. An LES study of flow across the junction of a mixing tee noted the significant reduction in temperature fluctuations with the presence of a porous interface \citep{lu2010large}. The averaged flow fields of the case with the porous junction also showed smoother velocity vectors towards the outlet. Similar dynamics could be at play with the presence of the stent (acting as the porous interface) obstructing the incoming distal flow (that could be likened to the branch of the mixing tee) yielding a smoother outlet flow with lower fluctuations.

In addition to this source of turbulence generation, the heel region of the stent-absent anastomosis was another location where high $TKE$ flow was emanating. The spatial extent and the magnitude of the $TKE$ increased at the maximum inlet flow time-point and was sustained in the deceleration phase of the inlet flow. Regions of high standard deviation of velocity have been noted at the anastomosis with a further increase at the heel during the deceleration phase (particularly with the intermediate angle) in the aforementioned LES study \citep{stella2019assessing}. Similar increases in disturbances at the anastomosis have been noted in other studies of AVF geometries \citep{Bozzetto2016, browne2015experimental} where it was postulated that the spiralling flow separated at the heel. Similar behaviour has been noted in the tortuous intracranial aneurysm geometries \citep{valen2014high} where disturbances tended to grow in the deceleration phase. 

Although the $TKE$ distribution is much lower in magnitude in the stented AVF, an increase in $TKE$ is noted at the heel of the anastomosis at the maximum time-point. The smaller recirculation region in the stented case, noted in \citet{gunasekera2021impact} and figure \ref{fig:velmag_amp}, led to lower fluctuations at the heel. The uneven wall surface, created by the stent struts, energised the boundary layer which in turn delays or negates flow separation \citep{lin1990investigation} and its ensuing turbulence generation. The increased $TKE$ noted at the heel of the stented case extended downstream along the stent at the maximum inlet flow time-point. However, at the deceleration time-point this extended region of $TKE$ starting from the heel of the AVF broke down into isolated smaller regions, again highlighting the increase in disturbances in the deceleration phase. Importantly, these fluctuations were constrained within the stent encapsulated region in the malapposed vein, with the stent acting as a buffer funnelling the turbulent fluctuations away from the vessel wall.

\subsection{Turbulence generation at the downstream edge of the stent implantation}
\label{sec:stent_end}

\begin{figure*}[h!]
\centerline{\includegraphics[width=15cm]{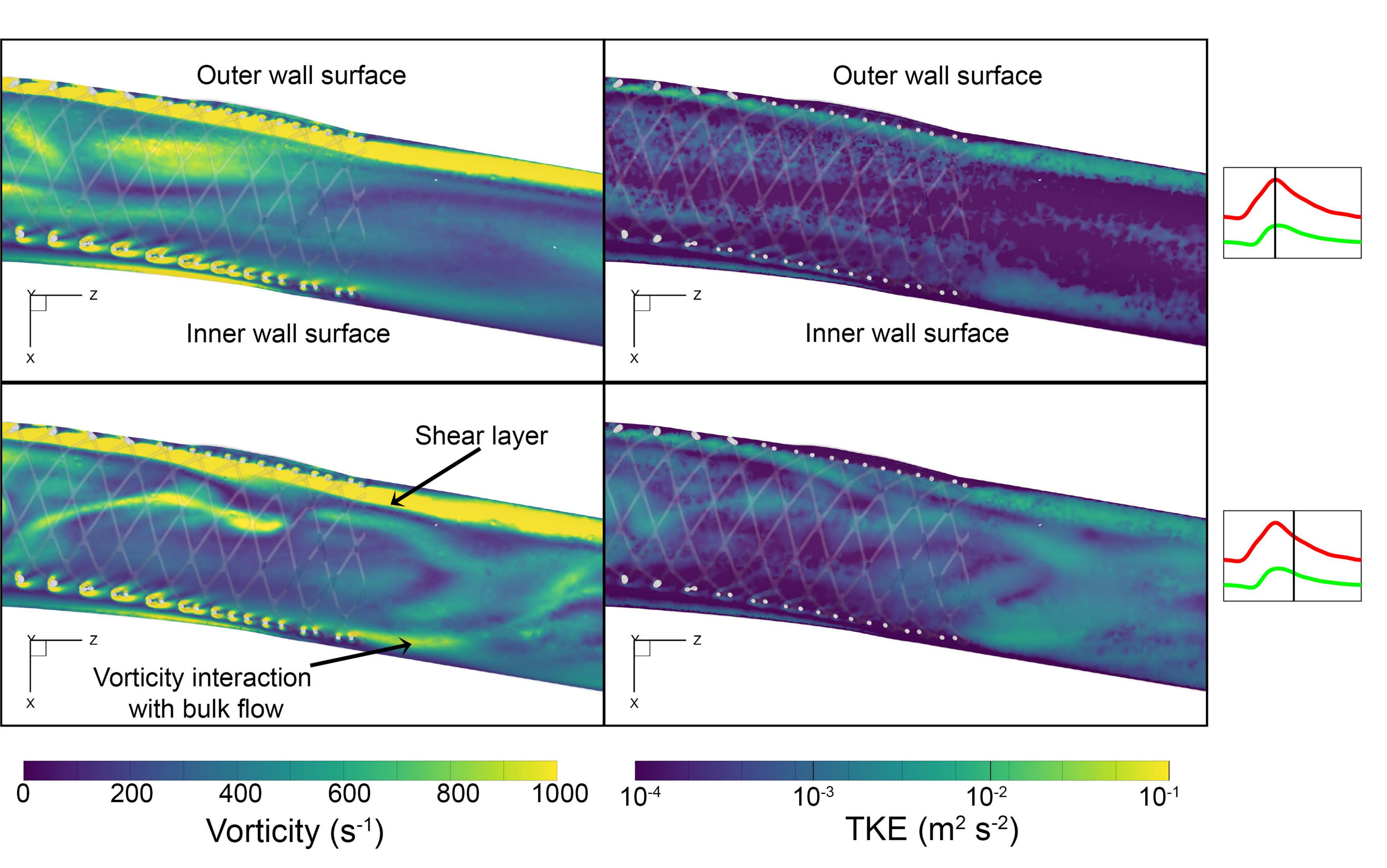}}
\caption{Contour plots of vorticity magnitude (left) and turbulent kinetic energy (right) on the central plane at the edge of the stent. The distributions have been plotted at the minimum, acceleration, maximum and deceleration inlet flow time-points for the stented case.}
\label{fig:stent_edge_plane_vort_tke}       
\end{figure*}

Although the noted $TKE$ is much lower in the stented case, there is an increase in $TKE$ magnitude (relative to other regions in the stented case) after the end of the stent. This increase in disturbance is seen most clearly in the deceleration phase of the inlet flow waveform. A study measuring shear stress of stented \textit{in vitro} vessels noted the presence of disturbances downstream of the stent implantation \citep{peacock1995flow}. The authors postulated that the struts could behave similar to a cylinder placed in a cross-flow where the oscillating vortices shed from the cylinder become unstable with flow over a certain Reynolds number. A recent PIV measurement of a stent-graft in an \textit{in vitro} compliant model noted a recirculation zone at the trailing edge that was postulated to be caused by the compliance mismatch between the graft and the artery \citep{yazdi2021vitro}. Vorticity magnitude was plotted across a central plane at the stent edge, as illustrated in figure \ref{fig:stent_edge_plane_vort_tke} to assess the span-wise vortical behaviour suggested in other studies. The magnitude (rather than a component) of the vorticity was used to avoid complications that arise with aligning the tortuous vessel with the principal axes.

The vortical behaviour is relatively low in the bulk flow of the vessel except at the deceleration phase where there is a high vorticity region coming off the heel of the anastomosis. Dean vortices that are created at the curved anastomosis breakdown in the deceleration phase due to the rotational energy being unsustainable with the decrease in momentum provided by the inlet flows. The magnitude of this feature reduces as the flow approaches the edge of the stent. More complex vortical behaviour is initiated in the bulk flow past the proximal edge of the stent (at the deceleration time-point). These complex regions are co-located with regions of non-zero $TKE$ as well.

Another striking feature is the high vorticity magnitude region at the stent struts. This behaviour is persistent throughout the cycle, however, the interaction between the vortical flow from adjacent stent struts is much higher at the outer wall surface (refer to the first row of figure \ref{fig:stent_edge_plane_vort_tke} for the inner/outer wall surface localities). The coalescing vortical flow structures amount to a shear layer forming and shielding the stent encapsulated flow. Studies on an idealised coronary artery geometry noted the coalescing of vortical regions with stent struts of shorter gaps as opposed to the creation of individual recirculation zones with longer strut gaps \citep{berry2000experimental}. However, in the current study, the stronger cohesion between adjacent vortical regions in the outer wall surface is due to the relatively higher velocity flow traversing the outer curved AVF anastomosis.

The shear layer is seen to continue past the stent edge at the outer wall surface, however, the shear layer at the inner wall surface is seen to interact with the aforementioned bulk flow complexities. This increase in magnitude and traversal of vortical flow towards the bulk flow region was noted in PIV studies of flow around a spanwise cylinder close to a wall \citep{he2017vortex}, where, at a critical gap width, the vortices created from the gap flow interact with the upper vortices. This vortical flow interaction leads to a significant increase in $TKE$ downstream of the stent edge during the deceleration phase. Figure \ref{fig:stent_edge_plane_vort_tke} illustrates the vortical flow interaction and the resulting $TKE$ increase on one plane, however, this behaviour is expected throughout the cross-section of the vessel circumscribed by the stent edge. 
 
\section{Impact of the turbulence generation on the wall shear stresses}
\label{sec:stent_wall}

To further illustrate the effects of the bulk flow features on the patient-specific vessel wall, a MATLAB script was developed to unwrap the vessel wall onto a 2D surface plot. This script utilised the vessel centerline node locations together with the vessel wall locations. A local centerline node direction vector can be calculated using a central difference of the two adjacent centerline node locations. Another direction vector, perpendicular to the local centerline direction vector, was created such that it pointed to the inner surface of the vessel from the centerline node. In the case of the outlet vein, the inner half of the vessel was the half that was closest to the proximal artery vessel, and vice-versa. The inner surface of the distal artery was also determined relative to the location of the proximal artery. Following these two steps provides a local axial vector and an auxiliary in-plane (vessel cross-section) vector to calculate distances away from the anastomosis and angles relative to the inner surface, respectively, at each node. 

\begin{figure*}[h!]
\centerline{\includegraphics[width=15cm]{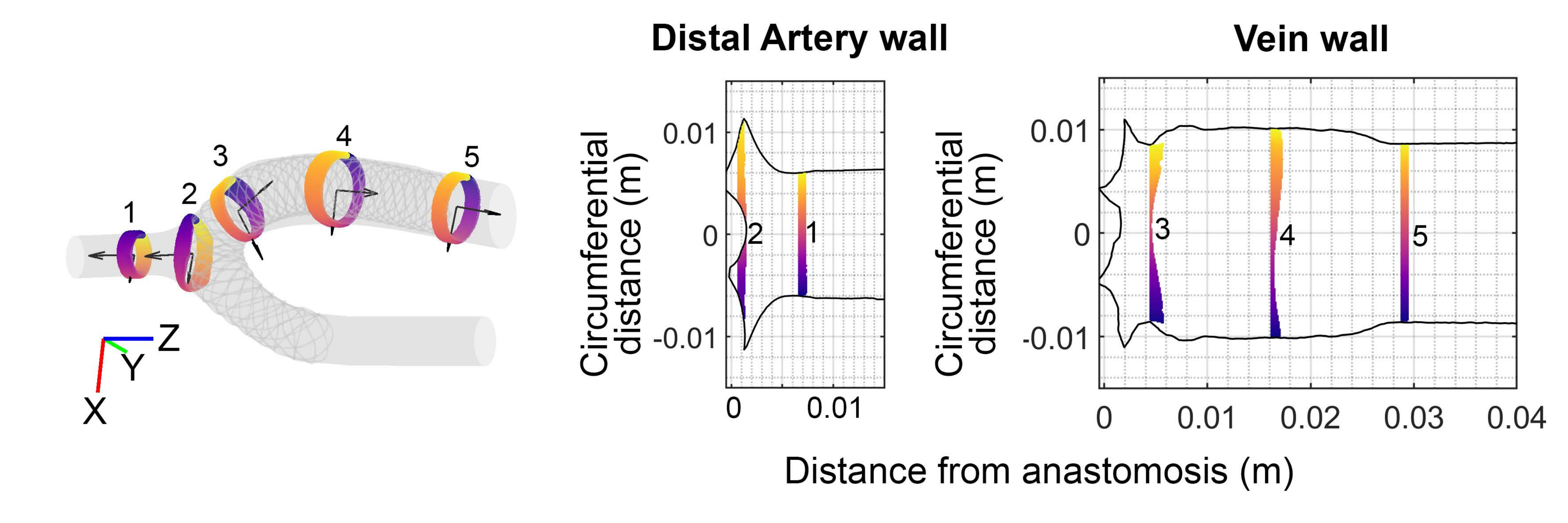}}
\caption{Example of unwrapping the 3D vessel wall locations onto a 2D space using local vectors constructed from centerline node locations. The colour bar of the cross-sectional slices assist with understanding the 3D location of the wall element on the corresponding 2D plots.}
\label{fig_stented_unravel}
\end{figure*}

The next step was to assign each wall element to its closest centerline node. Using the in-plane vector, an angle was measured for each of these assigned wall elements relative to the inner surface of the vessel. Assuming that the vessel cross-sections were close to circular, a circumferential distance of the wall element from the inner surface could be calculated using the aforementioned angle and the distance of the wall element from the centerline node. 

Following these steps yields an axial location along the vein and a circumferential location along the cross-section of the vessel for each wall element, thereby converting the prior 3D vessel wall surface to a 2D surface. The spacing of the centerline nodes was reduced to $0.1~mm$ increments to decrease unnatural steps in the plot. The clarity obtained by visualising the whole tortuous vessel on a plane outweighed the effects of the circularity assumption.

To quantify the impact of the bulk flow features on the vessel, both cycle-averaged and phase-averaged WSS metrics have been used. Distributions of time-averaged wall shear stress (TAWSS), defined by equation \ref{eq:tawss}, are used to note regions of low TAWSS which is inter-related to regions of near-wall low velocity. The Oscillatory Shear Index (OSI) captures the level of reversal of the WSS vector along the line of the dominant WSS vector \citep{he1996pulsatile} and is calculated using equation \ref{eq:osi}. Additionally, the TAWSS metric only provides a mean measure, while the OSI only captures the fluctuation along a single direction. Because of these deficiencies, additional WSS metrics were introduced, one of which is transverse wall shear stress (transWSS). This metric attempts to capture the multi-directional variation of the WSS vector across the cardiac cycle and is calculated by equation \ref{eq:transwss}.

\begin{equation}
\label{eq:tawss}
    TAWSS = \frac{1}{T}\int_{0}^{T} |\vv{WSS}| ~dt \\
\end{equation}

\begin{equation}
\label{eq:osi}
    OSI = 0.5\bigg(1-\frac{\frac{1}{T}|\int_{0}^{T}\vv{WSS} ~dt|}{TAWSS}\bigg)\\
\end{equation}

\begin{equation}
\label{eq:transwss}
    transWSS = \frac{1}{T}\int_{0}^{T}\bigg|\vv{WSS}\cdot \bigg(\vv{n}\times \frac{\int_{0}^{T}\vv{WSS} dt}{|\int_{0}^{T}\vv{WSS} dt|}\bigg)\bigg|dt
\end{equation}

In a similar fashion to the calculation of phase-averaged TKE, the phase-averaged turbulent fluctuations of WSS were also calculated at the maximum and deceleration time-points in the cycle. Using the fluctuations for each component of the WSS vector ($\vv{WSS}$), ${WSS'}_{x}=WSS_x-\overline{WSS_x}$, ${WSS'}_{y}=WSS_y-\overline{WSS_y}$, ${WSS'}_{z}=WSS_z-\overline{WSS_z}$, the magnitude of the WSS fluctuations were calculated as follows and shall be referred to as Turbulent Wall Shear Stress ($TWSS$). $TWSS$ values have been phase-averaged across 12 cycles to capture persistent distributions.

\begin{equation}
\label{eq:twss}
    TWSS = \sqrt{({WSS'}_{x}^{2}+{WSS'}_{y}^{2}+{WSS'}_{z}^{2})}
\end{equation}

\subsection{Behaviour of the WSS at the distal artery}

\begin{figure*}[h!] 
\centerline{\includegraphics[width=15cm]{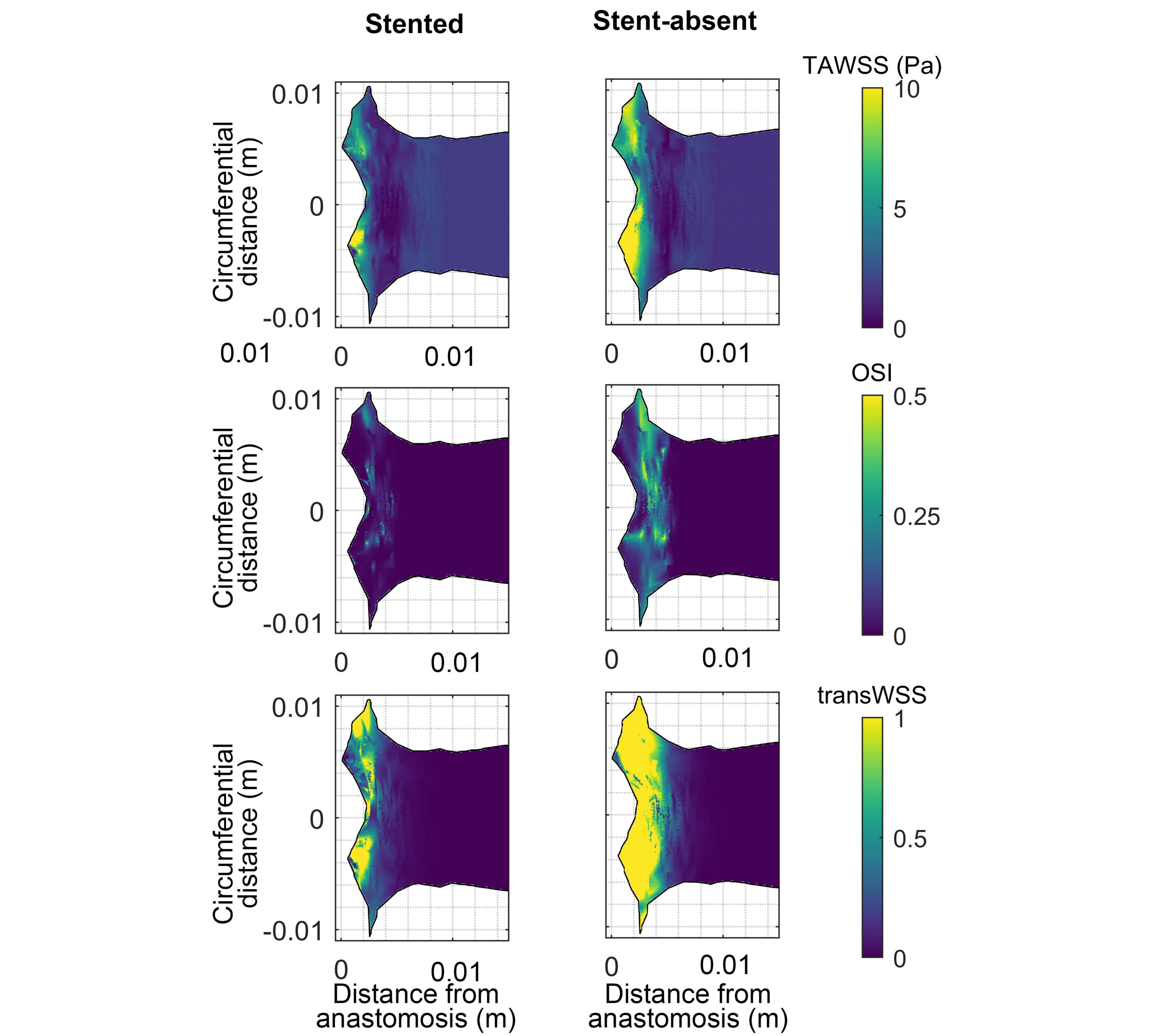}}
\caption{Wall shear stress metrics on vessel wall at the distal artery.}
\label{fig:da_wssmetrics}       
\end{figure*}

There is a clear low TAWSS region in the distal artery near the anastomosis for both the stented and stent-absent cases. This is to be expected as the collision of the two inlet flows creates a recirculation zone near the floor of the anastomosis. Since the trajectory of the coalesced flow is directed towards the vein and the high velocity flow moves towards the outer wall of the vessel (relating to the maximum circumferential distances in \cref{fig:da_wssmetrics}), the TAWSS starts increasing in the anastomosis. The increase is more sudden and persistent in the stent-absent case compared to the stented case. This is due to the stent implantation inhibiting the high velocity flow from being incident on large sections of the vessel wall.

In the stent-absent case, large regions of high OSI are noted in the same regions where low TAWSS was observed. Similar patterns are noted in the stented case as well, however, the magnitude and spread of these regions are very much subdued. The generation of high $TKE$ in the anastomosis of the stent-absent model could explain the high regions of oscillating wall shear stresses. However, transWSS is a better metric to assess multi-directional disturbances in the anastomosis. As expected, there are much larger regions of high transWSS noted in the stent-absent case when compared to that of the stented case. An important observation is that the high transWSS and OSI regions of the stent-absent AVF case are adjacent or coincident with the band of low TAWSS.

\subsection{Behaviour of the WSS in the juxta-anastomotic vein}

\begin{figure*}[h!] 
\centerline{\includegraphics[width=15cm]{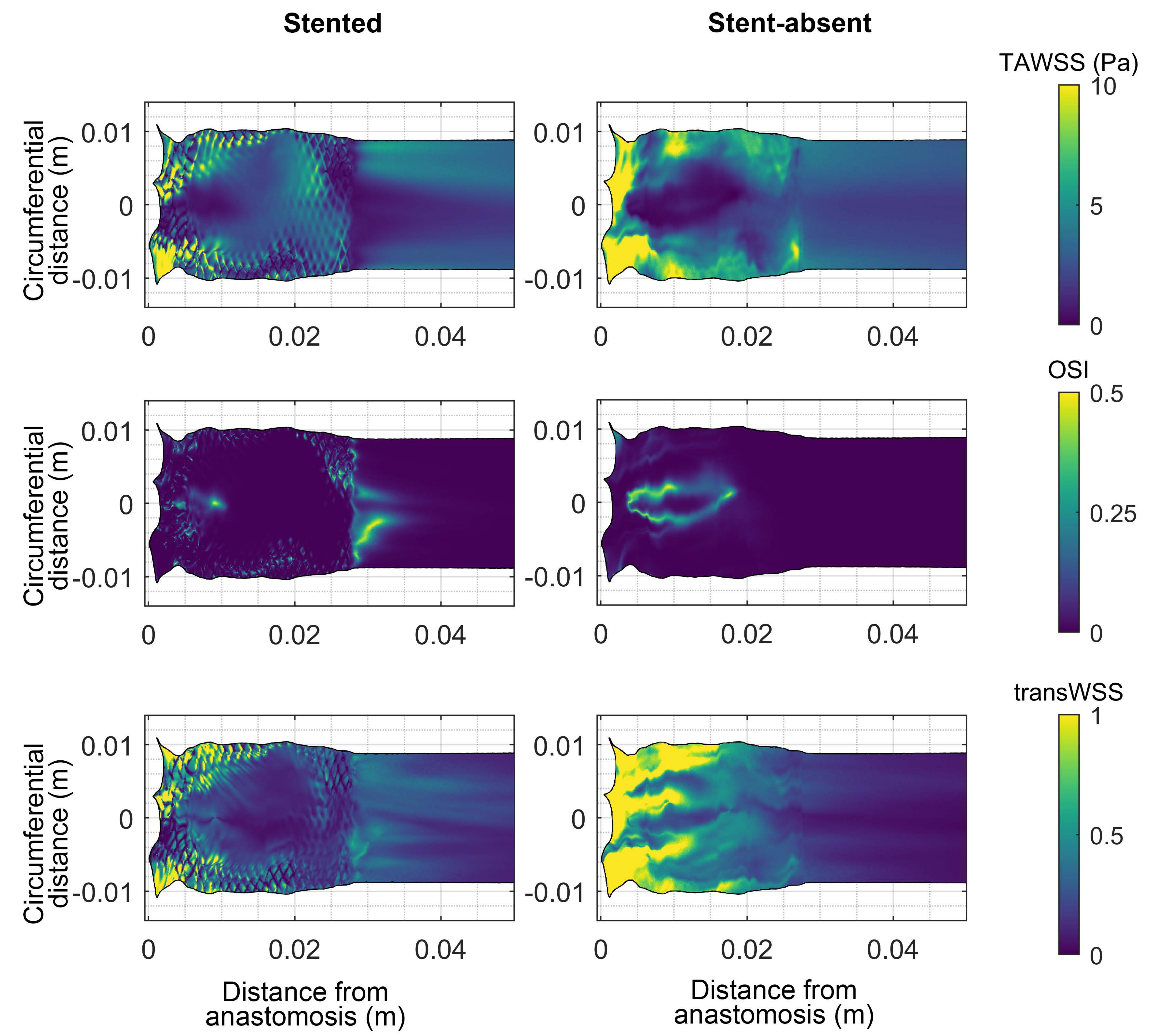}}
\caption{Wall shear stress metrics on vessel wall at the proximal vein.}
\label{fig:pv_wssmetrics}       
\end{figure*}

The distribution of the cycle-averaged WSS metrics: TAWSS, OSI, and transWSS, across the vein is illustrated in figure \ref{fig:pv_wssmetrics}. A clear difference between the stented and stent-absent cases is noted at the heel region of the vein. A large TAWSS accompanied by a large ring of high OSI is noted in the stent-absent case, however, the stent-absent case shows a much smaller region of high OSI. The funnelling capability of the porous stent is a major contributing factor to the decrease in OSI \citep{gunasekera2021impact}. The disturbances generated in the anastomosis are kept away from the vessel wall in the malapposed region of the vessel. The impact of the shorter recirculation zone results in a shorter region of low TAWSS in the stented case.

\begin{figure*}[h!] 
\centerline{\includegraphics[width=15cm]{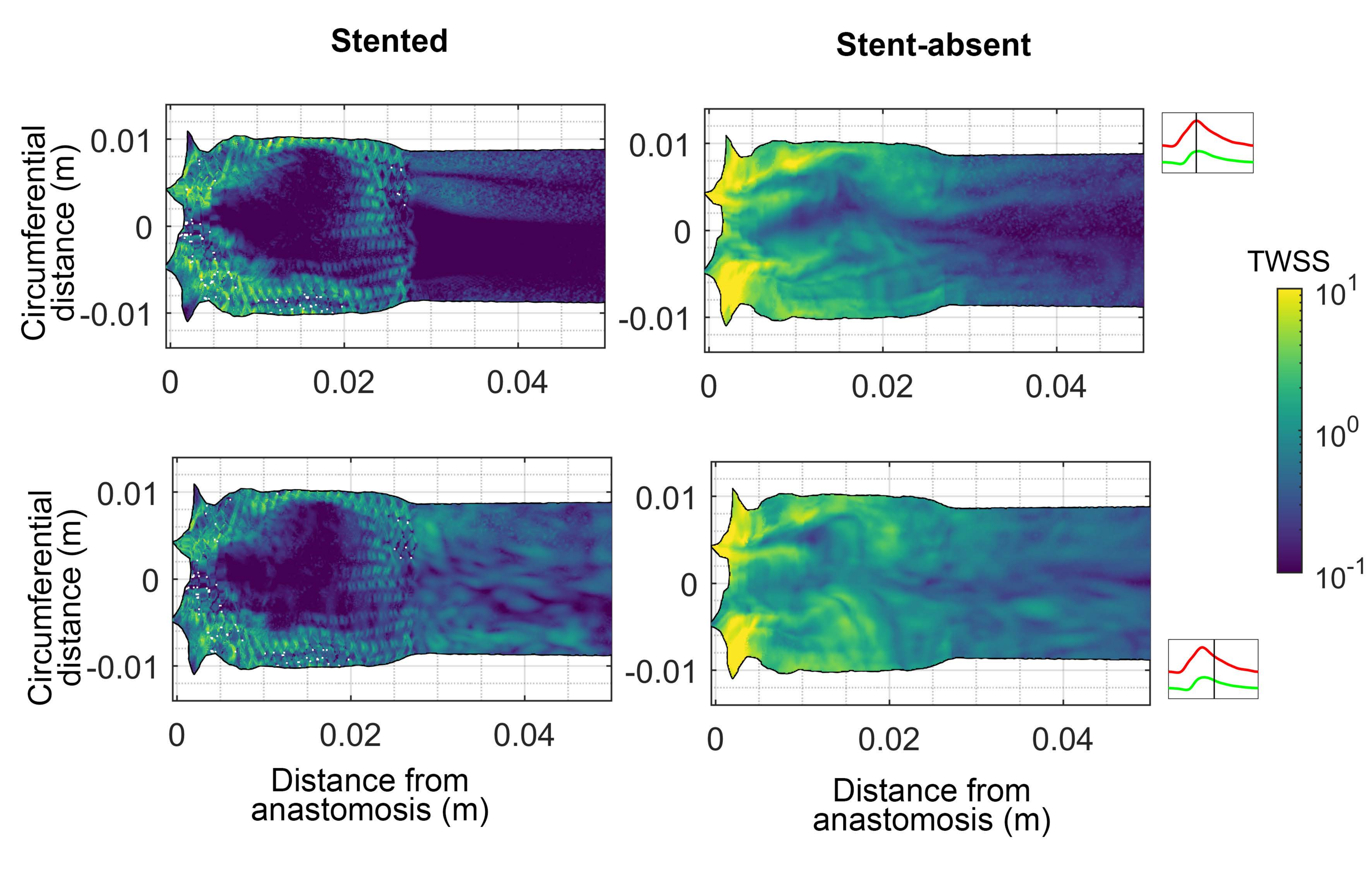}}
\caption{Turbulent Wall Shear Stress distribution at the maximum and deceleration time-points across the proximal vein.}
\label{fig:pv_twss_max_decel}       
\end{figure*}

There is a large distribution of high transWSS on either side of the recirculation zone in the stent-absent case, with the magnitude decreasing at the downstream edge of the recirculation zone. There is also a large distribution of high transWSS at the outer surface of the vein (relating to the maximum circumferential distances in \cref{fig:pv_wssmetrics,fig:pv_twss_max_decel}). The oscillatory flow behaviour that was initiated at the interface and conveyed along the outer bounds of the vein (as detailed with the temporal flow rate ratios of section \ref{sec:stent_recirc}) would lead to the vessel wall continually sensing a high and low velocity gradient across the cycle, with variations in the WSS vector direction. These flow oscillations were very much tempered in the stented case which resulted in a smaller distribution of the high transWSS. 

To better understand the impact of the $TKE$ flow behaviour, the distribution of the TWSS has been considered at the maximum and deceleration time-points in figure \ref{fig:pv_twss_max_decel}. A persistent high TWSS region is seen at the outer regions of the vessel wall near the anastomosis in the stent-absent case throughout the cycle. A possible contributor to these fluctuations in WSS is the $TKE$ generated at the interface of the two inlets. High amplitude oscillations were noted at the first three monitor points in the vein (refer to figure \ref{fig:fft}) and the variation in the velocity magnitude distribution is significant at the outer wall of the vein (refer to figure \ref{fig:velmag_amp}). The TWSS affirms that the cycle-to-cycle variations in the oscillatory flow also translate to cycle-to-cycle variations in the WSS. The TWSS is an order of magnitude lower near the heel of the anastomosis. Low TWSS is surrounded by a region of relatively higher TWSS, in a manner similar to that of the TAWSS and transWSS distributions. This behaviour is more apparent at the maximum time-point. The low velocity flow present at the heel would yield relatively lower magnitude fluctuations leading to lower fluctuations in the WSS as well. However, the variation of the extent of the recirculation zone leads to the high WSS fluctuations bordering the low velocity region.

The main behaviour seen in the stented case is that the high TWSS is concentrated to the stent strut regions. Studies have shown the presence of recirculation zones in the spaces between struts of apposed \citep{jimenez2009hemodynamically} and malapposed \citep{chen2017haemodynamic} stent segments. The disturbances caused in these micro-recirculation zones coupled with the high velocity flow traversing on the outer edge of the vessel yield significantly high levels of TWSS localised to the stent strut spaces. It is also clear that these dynamics localised around strut regions and effects of fluctuating bulk flow are not propagated to the vessel in the malapposed region where the TWSS magnitude is significantly low.

\subsection{WSS behaviour proximal to the stent implantation}

The WSS behaviour appears to be generally homogeneous in the stented case downstream of the heel until the flow reaches the stent edge. There is an increase of OSI and transWSS at the inner vessel wall (relating to a circumferential distance of zero in \cref{fig:pv_wssmetrics}). The increase in these WSS measures is co-located with a decrease in TAWSS which would be caused by the recirculating flow emanating from the malapposed stent edge. The decrease in shear stress and presence of recirculatory flow has also been noted in simulations of peripheral arteries with insufficient stent-graft apposition \citep{al2017hemodynamic}. The variation of the vortical flow emanating from the stent edge leads to the increase in the transWSS and OSI which spans an extent of approximately $7.5~mm$ downstream of the stent-edge.

The TWSS magnitude is low downstream of the stent-edge at the maximum time-point. However, there is a streak of high TWSS on the outer vessel wall caused by the higher velocity flow and the shear layer extending off the outer circumference of the stent-edge. This streak of high TWSS continually decreases in magnitude downstream of the vessel. The magnitude of this high TWSS feature is still approximately an order of magnitude lower than the regions surrounding the struts that are located on the outer walls and near the anastomosis. 

The TWSS distribution at the deceleration time-point is very different, with the whole vessel wall covered by clumps of relatively high TWSS regions. The dynamics of the interactions of the multiple vortices (detailed in \ref{sec:stent_end}) lead to a sudden increase in the $TKE$ near the wall that subsequently affects the near-wall velocity gradients as well. The increased fluctuating near-wall dynamics caused by the stent edge could lay claim to the build-up of IH downstream of stent implantations noted in aorta subjects \citep{barth1990flexible} and could potentially affect stented AVFs in a similar manner.

\section{Conclusion}
\label{sec:stent_conclusion}

Large eddy simulations of the flow within a patient-specific AVF, with and without a stent implantation, were conducted. Oscillatory behaviour was noted to be originating from the distal toe of the anastomosis of the AVF. The frequency peaks of the oscillatory behaviour were much larger in the stent-absent case. This behaviour was noted to continue throughout the inlet cardiac cycle, however, the subdued oscillatory flow in the stented case persisted in the deceleration phase of the inlet flow. Furthermore, a high $TKE$ region was consistently noted at the interface of the two inlet flows at all analysed time-points in the stent-absent case. This coincidence in findings confirms that the oscillatory flow behaviour is involved in the generation of turbulent flow in the anastomosis which is conveyed towards the vein. The levels of $TKE$ in the anastomosis of the stented case were far lower indicating that the presence of the stent at the location of the inlet flow interface could mitigate the flow disturbances.

An increase in $TKE$ levels was also noted at the heel of the stent-absent AVF anastomosis at the maximum and deceleration time-points. These flow features led to a ring of high WSS fluctuations within the cycle (OSI, transWSS) and across cycles (TWSS). The increase in $TKE$ was far less significant in the stented case and these disturbances were held within the stent encapsulated region. The stent circumvented these minor disturbances away from the vessel walls thereby resulting in a smaller region of adverse WSS behaviour at the heel of the AVF. However, there were localised regions of high WSS fluctuations surrounding the stent struts due to the generation of micro-recirculation regions. 

Although the fluctuating haemodynamics of the stented case were far less apparent than that of the stent-absent case, an increase in $TKE$ was noted at the downstream edge of the stent which was malapposed from the inner vessel wall. Complex vortical flow behaviour emanating from this malapposed stent edge acted as a catalyst in the increase in flow disturbances, particularly at the deceleration time-point of the inlet flow. These observations were mirrored with the TWSS measurements which also increased downstream of the stent edge. 

Therefore, the stent placement mitigated the initiation of $TKE$ and the malapposition at the heel shielded the vessel from the low levels of turbulent flow conveyed downstream from the anastomosis. However, the malapposition at the stent edge could potentially create adverse disease-causing flow conditions downstream of the implantation. This detailed investigation of the vascular procedure provides key understanding of the reasons behind the success of this treatment strategy from a fluid dynamics perspective.

\begin{acknowledgements}
Sanjiv Gunasekera and Olivia Ng were recipients of the Australian Government Research Training Program Scholarship during the course of this work and gratefully acknowledge this support.
\end{acknowledgements}


\bibliography{main_apssamp}

\end{document}